# Radio-Frequency Detection of Fabry-Pérot Interference and Quantum Capacitance in Long-Channel Three-Dimensional Dirac Semimetal $Cd_3As_2$ Nanowires

Sung Jin An,[1,¶] Jisu Kim,[1,2,¶] Myung-Chul Jung,[3,4,¶] Kidong Park,[1] Jeunghee Park,[5] Seung-Bo Shim,[6] Hakseong Kim,[6] Zhuo Bin Siu,[7] Mansoor B. A. Jalil,[7] Christian Schönenberger,[8] Nojoon Myoung,[3,4,*] Jungpil Seo,[2,*] and Minkyung Jung[1,9,10,*]

[1] DGIST Research Institute, DGIST, Daegu 42988, Korea

[2] Department of Physics and Chemistry, DGIST, Daegu 42988, Korea

[3] Department of Physics Education, Chosun University, Gwangju 61452, Korea

[4] Institute of Well-Aging Medicate & CSU G-LAMP Project Group, Gwangju 61452, Korea

[5] Department of Advanced Materials Chemistry, Korea University, Sejong 30019, Korea

[6] Quantum Technology Institute, Korea Research Institute of Standards and Science, 34113 Daejeon, South Korea

[7] Electrical and Computer Engineering, National University of Singapore, Singapore 117576, Singapore

[8] Department of Physics, University of Basel, Klingelbergstrasse 82, CH-4056 Basel, Switzerland

[9] Institute of next generation semiconductor technology (INST), DGIST, Daegu 42988, Korea

[10] Department of Interdisciplinary Engineering, DGIST, Daegu 42988, Korea

**ABSTRACT**

We demonstrate phase-coherent transport in suspended long-channel $Cd_3As_2$ nanowire devices using both direct current (DC) transport and radio-frequency (RF) reflectometry measurements. By integrating $Cd_3As_2$ nanowires with on-chip superconducting LC resonators, we achieve sensitive detection of both resistance and quantum capacitance variations. In a long-channel device ($L \sim 1.8$ μm), clear Fabry-Pérot (FP) interference patterns are observed in both DC and RF measurements, provide strong evidence for ballistic electron transport. RF reflectometry reveals gate-dependent modulations of the resonance frequency, arising from quantum capacitance oscillations induced by changes in the density of states and FP interference. These oscillations exhibit a quasi-periodic structure that closely correlates with the FP patterns in DC transport measurements. In another device of a $Cd_3As_2$ nanowire Josephson junction ($L \sim 730$ nm, superconducting Al contacts), FP interference patterns are too weak to be resolved in DC conductance but are detectable using RF reflectometry. These results demonstrate the high quality of our $Cd_3As_2$ nanowires and the versatility of RF reflectometry, establishing their potential for applications in topological quantum devices, such as Andreev qubits or gatemon architectures.

¶ *S.-J.A, J.K. M.-C.J. contributed equally to this work.*

[*]Corresponding authors. Email: nmyoung@chosun.ac.kr, jseo@dgist.ac.kr, minkyung.jung@dgist.ac.kr

**INTRODUCTION**

Dirac materials, such as topological insulators, have attracted much attention in recent years due to their chiral surface states characterized by the non-trivial π Berry phases.[1-7] In these materials, surface states are protected by extra symmetries, such as time-reversal symmetry or mirror symmetry, forming two-dimensional (2D) Dirac points.[1-4] It has been proposed that Dirac nodes, which are three-dimensional (3D) analogues to 2D Dirac points (i.e., $Cd_3As_2$ shown in the upper inset of Fig. 1(c)), can be present at certain $\vec{k}$ values in the Brillouin zone, protected by crystalline symmetry.[2,4-11] These 3D Dirac materials, termed Dirac semimetals (DSMs), can host pairs of topological charges known as Weyl fermions when time-reversal symmetry or inversion symmetry is broken, effectively forming massless Dirac fermions at low-energy excitations. Because the topological charges act as sources and sinks of Berry's phases, a non-trivial topological invariant can be defined in the momentum space, which guarantees the presence of peculiar Fermi arc states on the surface.

Among 3D Dirac/Weyl semimetals, $Cd_3As_2$ has emerged as a particularly promising platform for topological quantum devices due to its unique bulk and surface band structures, which give rise to numerous intriguing physical phenomena and quantum devices.[2,12-14] For example, $Cd_3As_2$ exhibits strong negative magnetoresistance (MR) due to chiral anomaly and large positive MR, depending on the magnetic and electric fields.[12,15-22] Furthermore, it demonstrates various exotic quantum transport properties, including a unique 3D quantum Hall effect of Fermi arcs,[23-28] Aharonov-Bohm oscillations,[29-30] π / 4π Josephson effects,[31-35] controllable p-n junctions,[36] and quantum dots.[37] For quantum device applications, ballistic transport in $Cd_3As_2$ is a key requirement. While the observation of Aharonov-Bohm oscillations provides evidence for ballistic transport in $Cd_3As_2$ nanostructures, Fabry-Pérot (FP) interference, additional evidence of ballistic transport has not yet been clearly demonstrated in these nanostructures. This absence can be attributed to two possible sources that disrupt phase-coherent transport.[39] First, disorder originating from crystallographic imperfections formed during nanowire growth can significantly impact transport properties; such imperfections can be identified and analyzed using transmission electron microscope. Second, contamination of the nanowire surface during device fabrication can introduce significant scattering. Electron transport in nanowires is highly sensitive to surface conditions, where contaminations such as water molecules adsorbed onto the surface or trapped at the interface between the nanowire and substrate can introduce localized charge traps and scattering centers.[39-41] To address this issue, clean FP oscillations are often observed in suspended nanowire structures, where the effects of surface contamination and substrate-induced disorder are minimized.[39-41]

In this work, we observe FP interference in long-channel $Cd_3As_2$ nanowires with lengths of up to 1.8 μm using direct-current (DC) and radio-frequency (RF) reflectometry measurement techniques. Suspended $Cd_3As_2$ nanowire devices are fabricated at the ends of superconducting LC resonators, enabling both DC and RF reflectometry measurements. The resonance frequency and magnitude shift with gate voltage, indicating changes in the capacitance and resistance of the device as the carrier density in the nanowire is tuned. By simultaneously measuring the DC conductance and RF reflected signal as a function of both source-drain and gate voltages, we observe clear checkerboard patterns characteristic of FP interference. From the shifts in the resonator frequency within this regime, we extract capacitance variations corresponding to the quantum capacitance of the nanowire, which exhibits quasi-periodic oscillations synchronized with the FP interference patterns. We further investigate FP interference in a $Cd_3As_2$ nanowire Josephson junction coupled to an LC resonator. In this device, FP interference is only observed in the RF reflected signal, demonstrating the enhanced sensitivity of RF resonator measurements compared to DC conductance measurements.

**EXPERIMENTAL METHODS**

The $Cd_3As_2$ nanowires used in this work are grown by chemical vapor deposition. Details of the growth process and nanowire characterization are provided in Ref. 36-38. To investigate the electron transport properties of $Cd_3As_2$ nanowire devices using DC and RF measurement techniques, we couple the $Cd_3As_2$ nanowire devices to superconducting LC resonators, which serve as impedance matching circuits. This configuration enables highly sensitive detection of gate-tunable changes in resistance and quantum capacitance within the $Cd_3As_2$ nanowire, offering a powerful platform for probing coherent transport phenomena.[42-45] To fabricate the coupled devices, we first realize the superconducting LC resonator, which consists of a superconducting inductor coil ($L$) and its parasitic capacitor ($C$), forming a resonant circuit incorporating the $Cd_3As_2$ devices.[45] The LC circuits are fabricated using a 100-nm-thick niobium (Nb) thin film sputtered onto a highly resistive Si substrate and patterned via electron beam lithography and reactive ion etching (RIE) in a mixed $SF_6$/Ar-gas atmosphere. Figure 1(a) shows an SEM image of a typical lithographically defined LC circuit. A large pad (indicated by the blue rectangle) in the middle of the inductor coil is used as a bonding pad to connect to a transmission line in the measurement setup. The other end of the inductor is grounded through a Cd$_3$As$_2$ nanowire device. We implement a suspended structure to achieve a clean Cd$_3$As$_2$ nanowire system by eliminating interface disorders between the nanowire and substrate.[36] As shown in Figs. 1(b) and (c), the end of the LC circuit (indicated by the red rectangular box in (a)) is gapped from the ground plane by a length of ~ 1.5 μm, and two Ti/Au bottom gates are formed in the recess area of the Nb gap. Subsequently, a Cd$_3$As$_2$ nanowire with a diameter of 100 nm is transferred across the bottom gates (Device A) using a micromanipulator. Finally, two Ti/Au (5/145 nm) contacts are defined after etching the native oxide on the nanowire surface with Ar plasma. The channel length between source and drain contacts is ~ 1.8 μm. The nanowire segments above the bottom gates are suspended and separated from the bottom gates by ~ 70 nm, enabling effective gating. All measurements in this work were performed at $T = 20$ mK.

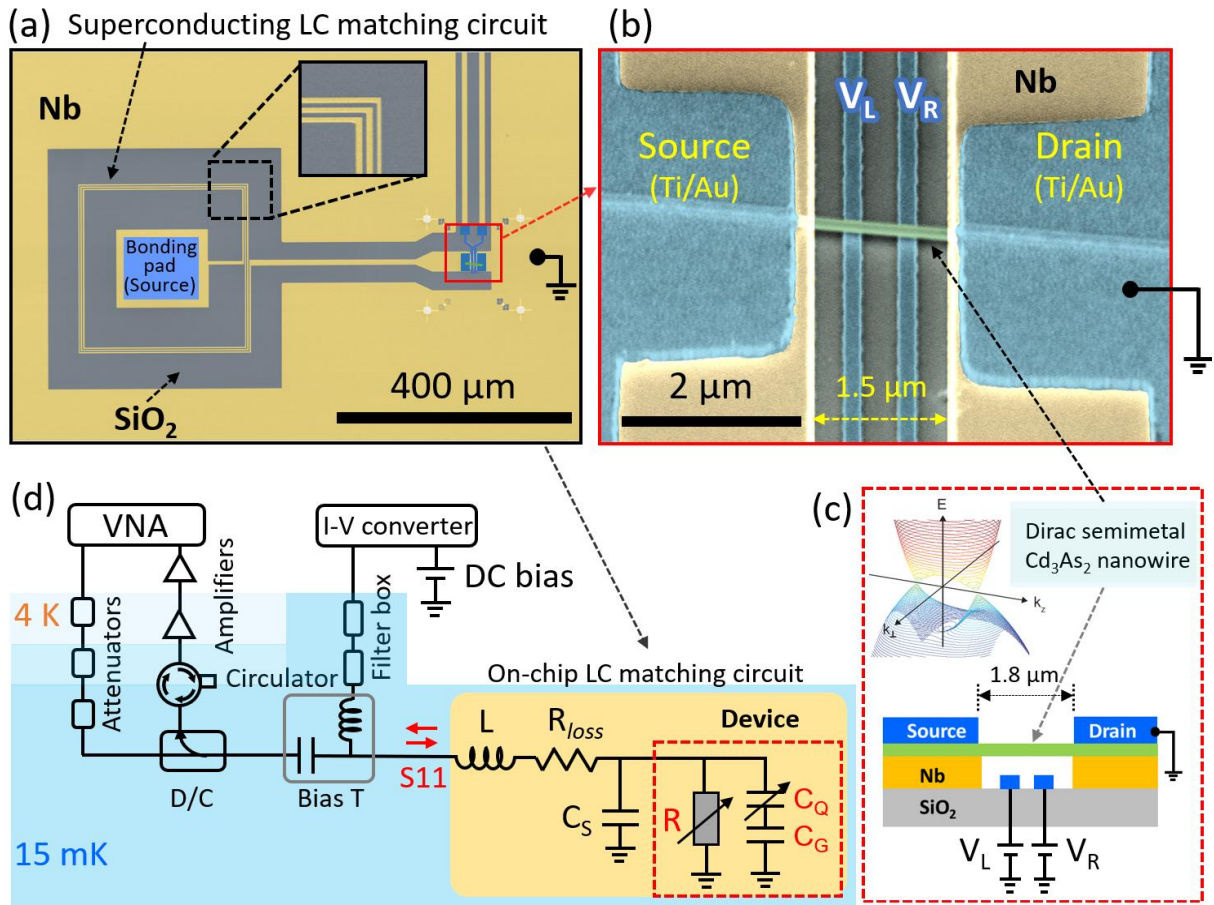

**Figure 1**. Cd$_3$As$_2$ nanowire device coupled to a superconducting LC resonator (impedance matching circuit). (a) SEM image of the device (Device A), showing a superconducting Nb on-chip coil inductor integrated with a suspended Cd$_3$As$_2$ nanowire structure fabricated on the same Nb film. DC measurements are performed through the bonding pad (blue) located at the center of the coil inductor. (b) Zoomed-in SEM image of the Cd$_3$As$_2$ nanowire device with Ti/Au source and drain contacts. The end of the LC circuit (red rectangular box in (a)) is separated from the ground plane by a gap of ~ 1.5 μm, and two Ti/Au bottom gates are formed in the recess area of the Nb gap. A Cd$_3$As$_2$ nanowire is transferred across the Nb gap and contacted with 150-nm-thick Ti/Au electrodes. The channel length between source and drain contacts is ~ 1.8 μm. (c) Schematic illustration of a suspended Cd$_3$As$_2$ nanowire device with two recessed bottom gates and the energy band structure of the Dirac semimetal Cd$_3$As$_2$. (d) Schematic of the measurement setup with the on-chip LC resonator coupled to the Cd$_3$As$_2$ nanowire device. RF reflectometry measurements are performed using a vector network analyzer (VNA) through a directional coupler, and DC measurements are conducted using an I-V converter through a bias-tee. The circuit model of the LC matching network and device components (orange box): the geometrical (gate) capacitor ($C_G$) and quantum capacitor ($C_Q$) of the device are connected in series to ground, while a parasitic capacitor ($C_S$) and the total capacitance ($C_T^{-1} = C_G^{-1} + C_Q^{-1}$) of the device are connected in parallel to ground.

The equivalent circuit diagram of the Cd$_3$As$_2$ nanowire coupled to the LC circuit is shown in the orange box of Fig. 1(d). The LC circuit comprises a superconducting inductor coil with an effective loss resistance ($R_l$) and a parasitic capacitance ($C_S$) to ground. The Cd$_3$As$_2$ device consists of a sample resistance ($R$) and a total capacitance ($C_T$). The total capacitance ($C_T$) of the device consists of a gate capacitance ($C_G$) in series with a quantum capacitance ($C_Q$), where the latter represents a correction term that depends exclusively on the band structure

$$C_T^{-1} = C_G^{-1} + C_Q^{-1} \text{ with } C_Q = e^2 \rho(E_F), \quad (1)$$

where $C_G = \varepsilon_r \varepsilon_0 A/d$, $A$ is the area of the parallel plate capacitor, $d$ is the dielectric thickness, $\varepsilon_r$ is the relative dielectric constant of the dielectric, $\varepsilon_0$ is the dielectric constant, and $\rho(E_F)$ is the 3D density of states in the materials.[42-45]

The total input impedance $Z_{in}$ of the system is given by

$$Z_{in} = j\omega L + R_l + \frac{R}{1 + j\omega R(C_S + C_T)}, \quad (2)$$

which yields the resonance frequency

$$f_r = \frac{1}{2\pi\sqrt{L(C_S + C_T)}} , \quad (3)$$

The effect of a change in $C_T$ is to shift $f_r$, which in turn alters the reflected signal. In reflectance measurements, the reflection coefficient ($\Gamma$) is given by

$$\Gamma \equiv \frac{Z_{in} - Z_0}{Z_{in} + Z_0} , \quad (4)$$

where $Z_0$ is the line impedance.

The entire measurement setup, illustrated in Fig. 1(d), follows that of Ref. 45. For the reflectometry measurements, a microwave signal with frequency $f_C$ is injected into the LC resonator input through a directional coupler. The reflected signal is then amplified by low-temperature and room-temperature amplifiers. We measure the reflection coefficient of the resonant circuit using a network analyzer while simultaneously performing DC transport measurements using a bias tee to apply a bias voltage at the same port.

**RESULTS AND DISCUSSION**

Figure 2(a) shows the differential conductance $G$ measured as a function of two bottom gates, $V_L$ and $V_R$. As $Cd_3As_2$ is a 3D analogue of graphene (see the energy band structure in the inset of Fig. 1(c)), we can modulate both the carrier type and density across gate segments using $V_L$ and $V_R$ to form p-n junctions. However, within our measurement range, we observe only the n-n transport regime of the device. Figure 2(b) shows the differential conductance $G$ measured along the dashed line in Fig. 2(a). As the gate voltage decreases, the conductance minimum indicative of the Dirac nodes is not observed in this voltage range because the Dirac nodes are shifted far away from $V_L = V_R = 0$ V, likely due to an unintentional doping effect in the nanowire. The conductance exhibits rapid conductance oscillations superimposed on a slowly varying conductance, characteristic of FP interference in the ballistic transport regime, which we analyze further in Fig. 3. Simultaneously with the DC transport measurements, we perform the RF reflectometry measurements using the network analyzer with an incident power of –70 dBm applied to the resonant circuit. As shown in Fig. 2(c), the LC circuit exhibits a resonance frequency near 4.557 GHz, with both the frequency and magnitude modulated by $V_{\text{Diag}}$ ($V_L = V_R$). These modulations reflect changes in the capacitance and resistance of the device, respectively. Figure 2(d) presents a 2D map of the RF reflected signal measured as a function of $V_L$ and $V_R$ by setting the frequency at $f = 4.557$ GHz. The RF reflected signal shows behavior nearly identical to the DC conductance map shown in Fig. 2(a).

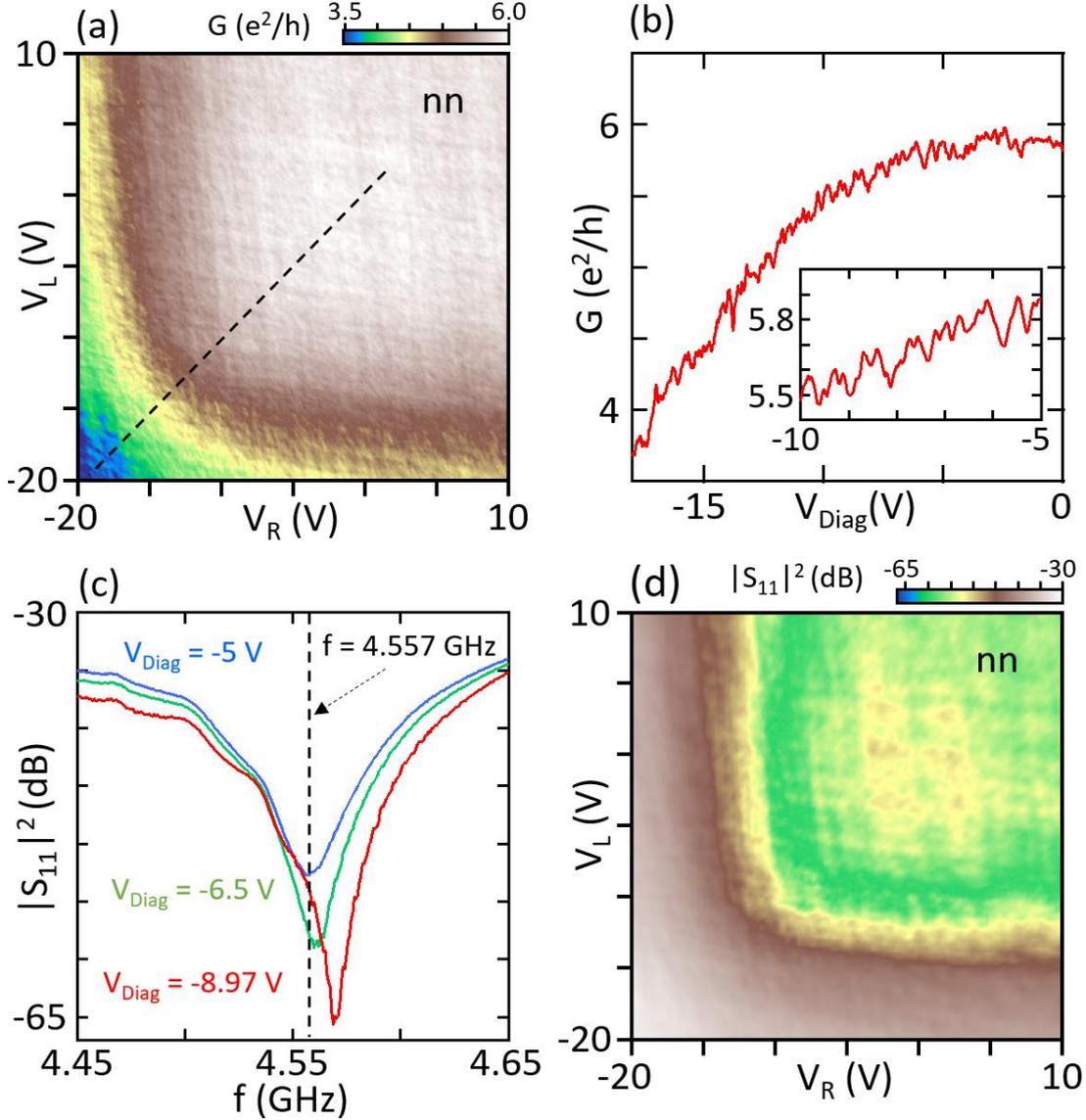

**Figure 2.** (a) Two-dimensional (2D) map of differential conductance measured in the n-n regime as a function of left ($V_L$) and right ($V_R$) gate voltages. (b) Line cut of the differential conductance along the dashed line in (a), plotted as a function of diagonal voltage $V_{Diag}$ ($V_L = V_R$), showing conductance oscillations attributed to FP interference. Inset: Magnified view of the oscillations between $V_{Diag} = -10$ V and $-5$ V. (c) RF Reflection coefficient ($|S_{11}|^2$) of the LC circuit measured as a function of frequency for different gate voltages, demonstrating shifts in resonance frequency (caused by changes in device capacitance) and magnitude (due to resistance variations). (d) 2D map of the reflection coefficient measured as a function of $V_L$ and $V_R$ at $f = 4.557$ GHz (marked by the black dashed line in (c)).

We investigate the origin of the conductance oscillations observed in Fig. 2(a) and (b). As shown in Fig. 3, we simultaneously measure the DC differential conductance (Fig. 3(a)) and the RF reflected signal (Fig. 3(b)) as a function of $V_{SD}$ and $V_{Diag}$ along the black dashed line in Fig. 2(a) (a diagonal sweep in the n-n regime). Quasi-periodic oscillations, which appear as checkerboard patterns in both the conductance and RF reflected signal, are clearly visible as the gate voltage changes (The full scale of 2D plot over the entire measurement range is provided in Supporting information Fig. S6). These oscillations are attributed to FP interference caused by electron waves bouncing back and forth between interfaces within the $Cd_3As_2$ nanowire channel. FP interference arises from the wave-like behavior of electrons confined in the nanowire. When electrons travel between two interfaces, such as the metal contacts at both ends of the nanowire, partial reflections occur at these boundaries. Constructive or destructive interference of these reflected electron waves results in conductance oscillations. The condition for constructive interference is given by $kL = m\pi$, where $k$ is the electron wavevector, $L$ is the FP cavity length, and $m$ is an integer representing the interference mode. Assuming a linear dispersion $\varepsilon(k) = \hbar v_F k$ with the Fermi velocity $v_F = 0.89 \times 10^6$ m/s of $Cd_3As_2$, this yields a single particle excitation energy $\Delta E = eV_{SD} = \varepsilon(k_{n+1}) - \varepsilon(k_n) = \hbar v_F \pi / L$.[39-41]

As shown in Fig. 3, the FP interference appears as checkerboard-like oscillations in the differential conductance. However, deviations from this regular pattern are evident, particularly near $V_{Diag} \approx -4.5$ to $-5.0$ V and again around $-2$ V. These irregular features are more clearly revealed in the fast Fourier transform (FFT) analysis presented in Fig. 3(c). A plausible scenario for this irregularity is the involvement of multiple 1D subbands in electron transport, as previously reported in the study by Kretinin et al. on suspended InAs nanowires (Ref. 39). In quasi-1D systems like $Cd_3As_2$ nanowires with diameters of ~ 100 nm, transverse confinement can result in the occupation of several 1D subbands, particularly under higher gate voltages. The number of transport channels in the $Cd_3As_2$ nanowire can be estimated as $n \approx 2W/\lambda_F$,[46] where $W$ is the nanowire width and $\lambda_F$ is the Fermi wavelength of $Cd_3As_2$ (~ 42 nm). Assuming a rectangular cross-section with $W \approx 100$ nm (corresponding to a cylindrical nanowire of diameter ~ 100 nm), we obtain $n \approx 4.7$, indicating that typically 4 – 5 channels can participate in transport. The experimentally measured maximum conductance, however, is ~ $6e^2/h$. This small discrepancy likely arises from the simplicity of the model, which does not account for contact resistance, residual scattering, and other experimental factors. In addition, in nanowire geometries, finite-size effects gap out the bulk states, whereas topological surface states persist inside this gap.[47] As a result, electron transport is dominated by surface states. Because their dispersion is effectively one-dimensional, the transport characteristics closely resemble those of semiconducting nanowires, and this supports the use of a multi-subband 1D model to simulate mode mixing in $Cd_3As_2$ nanowires.[47] When several subbands contribute simultaneously, each with its own Fermi wavevector and phase accumulation, the resulting interference conditions vary among subbands. This can lead to superpositions of FP oscillations with different periodicities and amplitudes, producing complex, nonuniform patterns in the conductance and the RF reflected signal.

Simple theoretical modeling (Supporting information), shown in Figs. 3(d) and (e), captures this behavior by simulating transmission through multiple 1D subbands.[39] The simulated conductance, obtained by summing the independent transmission probabilities of each subband, exhibits quasi-periodic oscillations superimposed on a slowly varying background. In regions where only the lowest subband is occupied, the FP oscillations appear more regular, as the constructive interference condition $kL = m\pi$ is satisfied uniformly (Supporting information Figure S4). However, as higher subbands become populated, the total conductance exhibits beating patterns and irregular FP modulation (Supporting information Figure S5). These arise from the superposition of phase-coherent contributions with different wavevectors and interference conditions, consistent with the multimode Landauer–Büttiker formalism. Thus, the observed irregular FP patterns likely reflect a transition from single-mode to multimode coherent transport within the $Cd_3As_2$ nanowire.

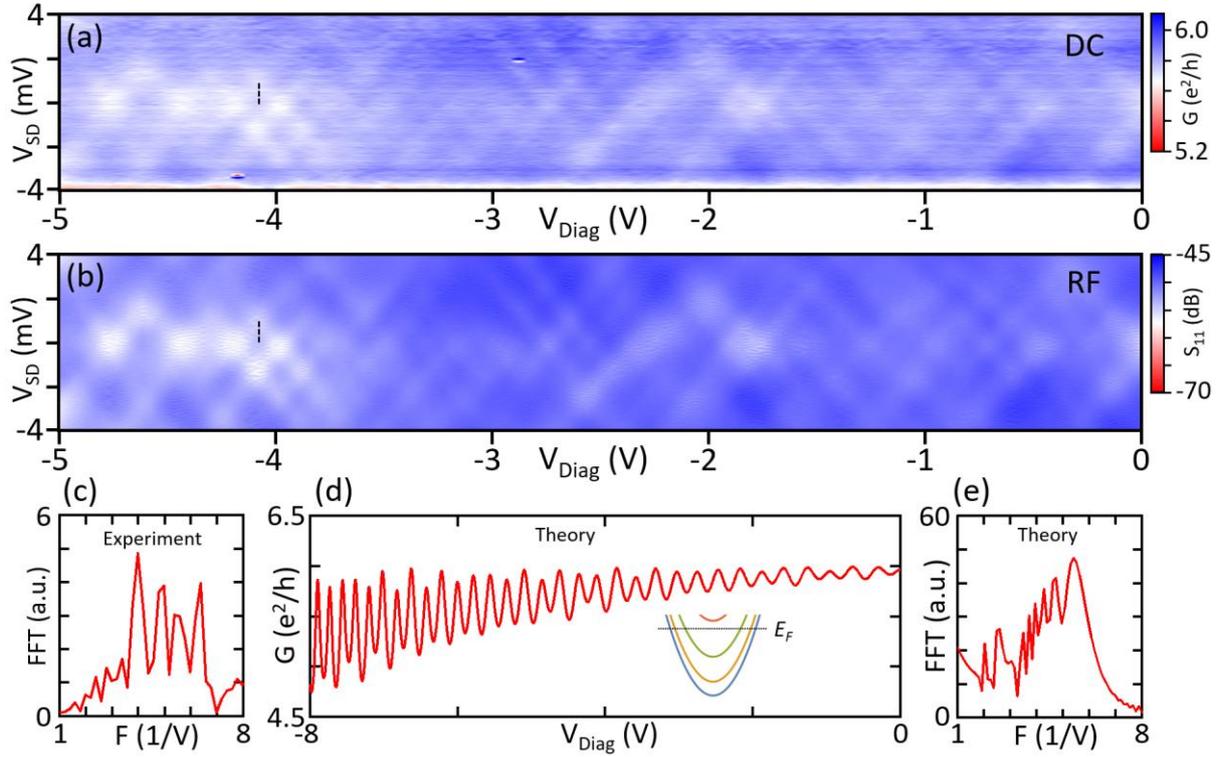

**Figure 3.** (a) Differential conductance and (b) RF reflection coefficient ($|S_{11}|^2$) measured along the dashed line in Fig. 2(a), plotted as a function of $V_{SD}$ and $V_{Diag}$. Checkerboard patterns arising from Fabry-Pérot (FP) interference are observed in both measurements, with a characteristic bias scale (highlighted by the dashed lines in (a) and (b)) of $V_{SD} \approx 1 - 1.1$ mV. The extracted FP cavity length is $L \approx 1.64 - 1.74$ μm, consistent with the physical nanowire length. Deviations from the regular FP interference pattern, particularly near $V_{Diag} \approx -4.5$ to $-5.0$ V and $-2$ V, can be attributed to multimode electron transport involving multiple 1D subbands. (c) Fast Fourier transform (FFT) of the experimental conductance data in (a), taking along $V_{Diag}$ at a fixed $V_{SD}$ = 0 V, revealing multiple frequency components. (d) Simulated conductance as a function of gate voltage ($V_{Diag}$) based on transport through three occupied 1D subbands. (Inset) Schematic illustration of the energy dispersion for the three subbands and the Fermi level $E_F$. (e) FFT of the simulated conductance from (d), showing multiple spectral components arising from multimode transport. The observed irregularity, and beating in the interference patterns can originate from the superposition of oscillations from distinct subbands, consistent with the multimode Landauer-Büttiker model.

In the other region of the gate voltage, the FP interference shows relatively uniform checkerboard patterns. From the characteristic bias $V_{SD} \approx 1 – 1.1$ mV, we extract an FP cavity length of $L \approx 1.64 – 1.74$ µm, which closely matches the physical length of the nanowire between the source and drain contacts. The observation of clear FP interference over a channel length exceeding 1.6 µm indicates robust phase-coherent electron transport, with coherence lengths that appear to be longer than those typically reported for other semiconductor nanowires such as InAs, InSb and $Bi_2Se_3$.[39,48-52] This result highlights the exceptional quality and low disorder of our fabricated $Cd_3As_2$ nanostructures. Despite the DC conductance and the RF reflectance data showing similar overall trends, the RF reflected signal exhibits slightly enhanced sensitivity to fine features in the interference patterns. Since RF measurements are less susceptible to low-frequency noise sources, they offer a higher signal-to-noise ratio. This improved sensitivity makes RF reflectometry particularly effective for detecting quantum interference patterns and for revealing subtle electron transport phenomena that are often obscured in conventional DC measurements.

Figure 4(a) shows a 2D map of the reflected signal measured as a function of frequency and $V_{Diag}$ along the black dashed line in Fig. 2(a). Clear gate-voltage ($V_{Diag}$) dependence is observed in both the resonance frequency and magnitude, reflecting changes in the capacitance and resistance of the $Cd_3As_2$ nanowire, respectively. The resonance frequency exhibits two distinct regimes: one is a nearly constant frequency region from $V_{Diag} \approx –2$ V to 0 V, and the other is a regime with pronounced frequency shifts from $V_{Diag} \approx –2$ V to $–10$ V. This behavior indicates a transition from gate capacitance-dominated transport to quantum capacitance-dominated transport, as discussed in Fig. 1. Small rapid resonance frequency oscillations are superimposed on the slow-changing resonance frequency shift, originating from changes in the density of states in the $Cd_3As_2$ nanowire.

To quantitatively analyze the rapid oscillations in resonance frequency, we extract the total capacitance change by fitting the resonance frequency data using the circuit model described by Eqs. (2) and (4). Figure 4(b) presents the extracted capacitance (blue curve, left axis) and the corresponding conductance (red curve, right axis) as a function of $V_{Diag}$. The fitted circuit parameters yield $R_l = 1.7$ Ω, $L = 16.37$ nH, and $C = 74.05 \pm \alpha$ fF, where α represents the contribution of the quantum capacitance arising from the density of states. Two distinct regimes appear depending on the relative magnitudes of quantum capacitance ($C_Q$) and the gate capacitance ($C_G$). For $V_{Diag} \gtrsim –2.0$ V, where $C_Q \gtrsim C_G$, the total capacitance ($C_T$) remains nearly constant, suggesting that the gate capacitance dominates, as expected from the series capacitance model (Eq. 1). In contrast, for $V_{Diag} \lesssim -2.0$ V, where $C_Q \lesssim C_G$, we observe an increasing $\Delta C_T$ accompanied by pronounced oscillations that correlate closely with conductance variations. Figure 4(c) and (d) show magnified 2D plots of the FP interference pattern from Fig. 3(b) and the corresponding quantum capacitance modulation measured between –5 V and –3.5 V, respectively. The quasi-periodic peaks and dips in capacitance align with the FP interference pattern, as indicated by the orange dashed lines. This clear correspondence suggests that the observed capacitance oscillations result from constructive and destructive interference of electron waves in the ballistic transport regime. In this voltage range, the change in quantum capacitance is estimated to be $\Delta C \approx 10 - 40$ aF.

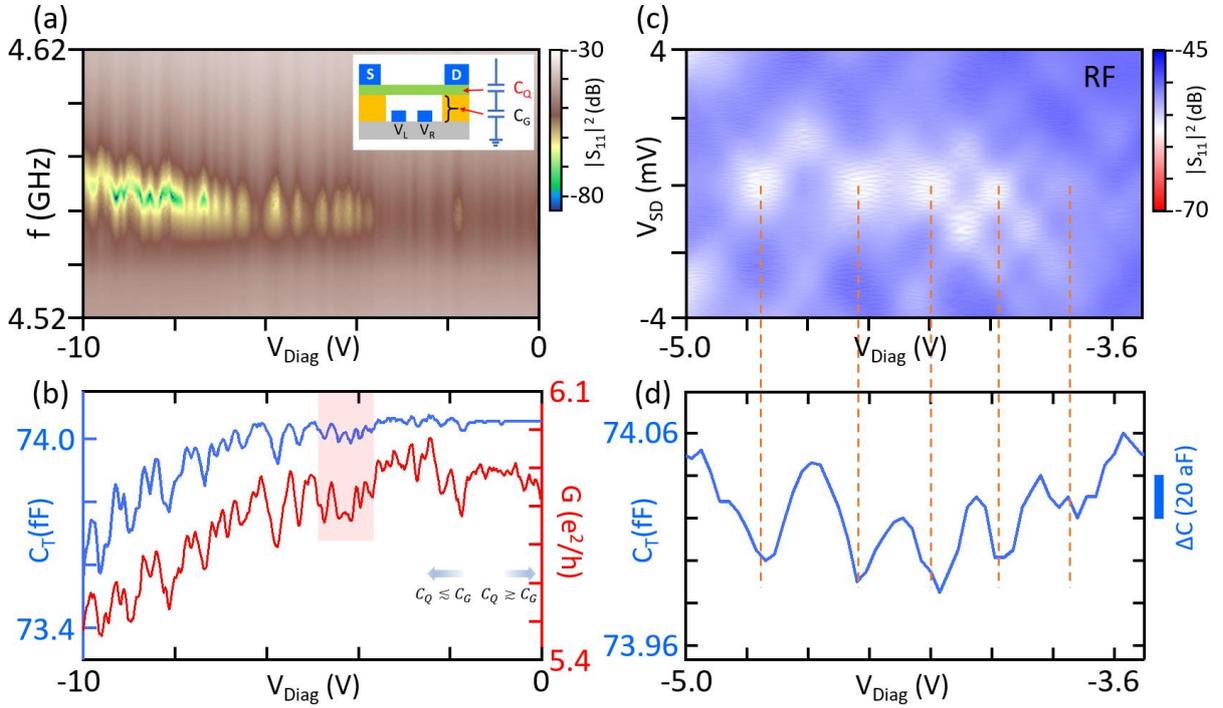

**Figure 4.** (a) Reflected power of the LC circuit near the resonance frequency as a function of resonance frequency $f$ and gate voltage $V_{Diag}$. A resonance frequency shift is observed as $V_{Diag}$ decreases. The inset shows the equivalent circuit model, including the quantum capacitance ($C_Q$) and gate capacitance ($C_G$). (b) Extracted total capacitance (blue, left axis) obtained by fitting the resonance frequency with Eq. 2 and 4, and conductance (red, right axis) versus $V_{Diag}$. Two distinct regimes are indicated based on the relative magnitude of the quantum ($C_Q$) and gate ($C_G$) capacitances, as indicated by colored arrows. (c) Magnified map of the RF reflected signal from Fig. 2(b), showing the FP interference pattern, and (d) corresponding quantum capacitance oscillations in the region $V_{Diag} \approx -5$ V to $-3.6$ V (highlighted by the pink shaded area in Fig. 4(b)). Orange dashed lines indicate the correlation between the interference patterns

and capacitance oscillations.

We further investigate FP interference in a suspended $Cd_3As_2$ nanowire Josephson junction device (Device B) with a channel length of 730 nm.[53,54] As shown in Figs. 5(a) and (b), the device is fabricated similarly to Device A. However, in this case, we use a single bottom gate within a Nb gap instead of two bottom gates, and Ti/Al source and drain contacts are employed to form the Josephson junction. This Josephson junction device is also integrated with a superconducting LC resonator, similarly to Device A. We perform similar DC transport and RF reflectometry measurements on Device B. Figures 5(c)-(f) display the DC conductance and the RF reflected signal maps measured as a function of $V_{SD}$ and $V_G$ at temperatures of 3.6 K and 15 mK. At 3.6 K, which is well above the superconducting critical temperature of Al, the device does not exhibit FP interference or a superconducting Josephson effect in either RF or DC measurements, as shown in Figs. 5(c) and (d). When the temperature is lowered to 15 mK, the device exhibits a conductance peak around $V_{SD} \approx 0$ V in both DC and RF measurements, due to the superconducting branch as the $Cd_3As_2$ nanowire becomes superconducting. Because Device B is integrated with an LC resonator, direct four-terminal measurements of the Josephson effect are not possible. To examine superconducting Josephson effect at a comparable channel length, we fabricate and characterize a separate $Cd_3As_2$ nanowire junction (~600 nm) using a four-terminal configuration (Device C, see Fig. S7 in Supporting Information). In this device, clear supercurrent branches are observed around zero current bias, confirming proximity-induced superconductivity in $Cd_3As_2$ nanowires at this length scale.

Although the channel length of Device B is approximately 2.5 times shorter than that of Device A, FP interference is not visible in the DC conductance, implying that Device B exhibits more diffusive transport. However, as shown in Fig. 5(f), the differential RF reflected signal reveals clear FP interference as a function of both $V_{SD}$ and $V_G$. This result indicates that the sensitivity of the LC resonator is sufficient to detect FP oscillations, which are not measurable in the DC conductance. The estimated FP cavity length is $L = hv_F/V_C \approx 550$ nm, which corresponds to the geometrical separation of the two Nb supporting structures. As illustrated in Fig. 5(g), we attribute the FP cavity formation to the interface between the suspended and supported segments of the nanowire, where a mismatch in the Fermi level induces partial reflection and standing wave formation.

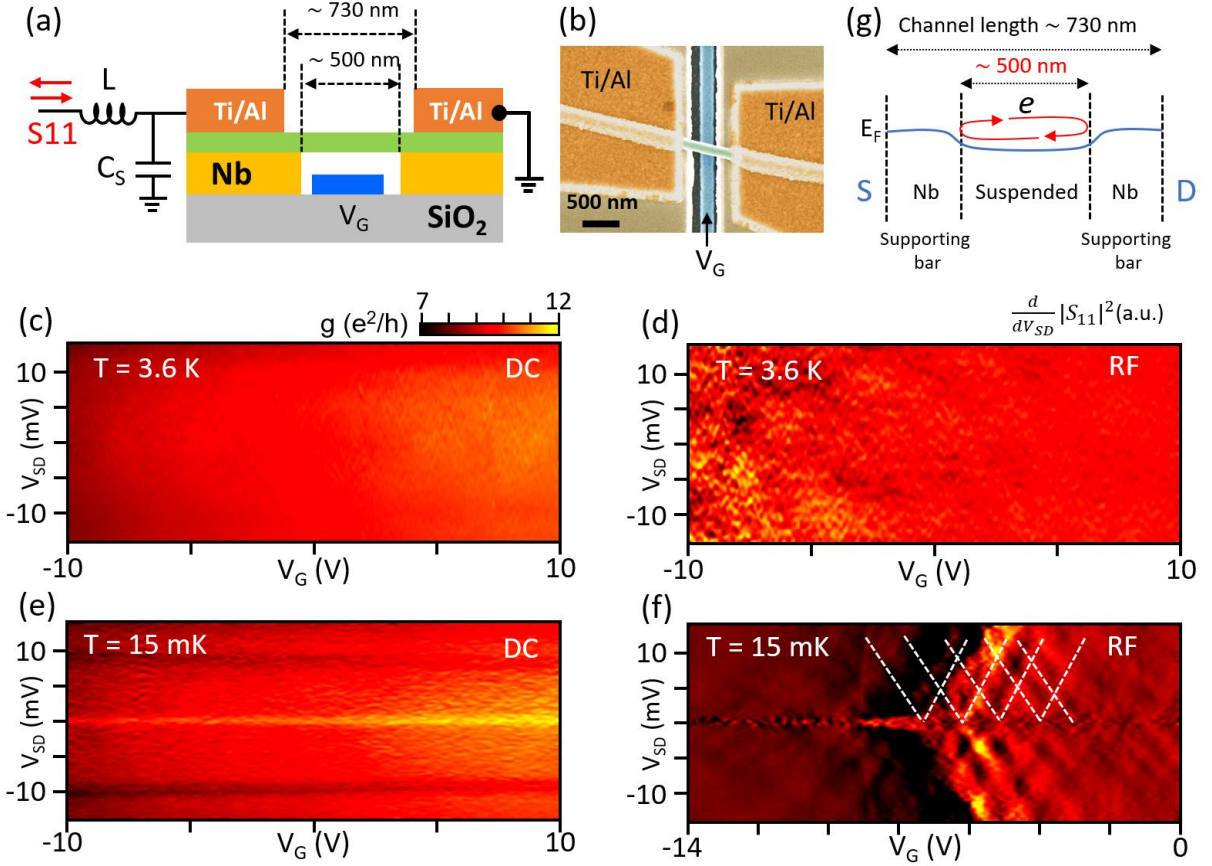

**Figure 5.** (a) Schematic of a suspended Cd$_3$As$_2$ nanowire Josephson junction device with a recessed single bottom gate coupled to a superconducting LC resonator (Device B). The total channel length and suspended segment of the nanowire are ~ 730 and ~ 500 nm, respectively. (b) SEM image of a suspended Cd$_3$As$_2$ nanowire Josephson junction. (c) DC conductance and (d) Differential reflection coefficient measurements at $T$ = 3.6 K. (e) DC conductance and (f) differential reflection coefficient measurements at $T$ = 15 mK. The white dashed lines are guides for the eye, highlighting the checkerboard pattern of FP oscillations at high bias. (g) Illustration of the FP interference cavity in the Josephson junction device (not to scale).

**CONCLUSION**

In conclusion, using both DC and RF reflectometry measurement techniques, we observed FP interference and quantum capacitance in 3D Dirac semimetal Cd$_3$As$_2$ nanowire devices. For simultaneous measurements of the DC conductance and RF reflected signal, Cd$_3$As$_2$ nanowires with a suspended structure were fabricated at the end of on-chip superconducting LC resonators. The device exhibits clear FP interference in both the DC conductance and RF reflected signal, a manifestation of ballistic transport in suspended long-channel $Cd_3As_2$ nanowires up to 1.8 μm. In our RF reflectometry measurements, the resonance frequency of the LC resonators is modulated as a function of gate voltage, attributable to quantum capacitance changes within the nanowire. We observed quasi-periodic rapid quantum capacitance oscillations superimposed on a slowly varying quantum capacitance, extracted from the resonance frequency fitting. The periodicity of the measured FP interference is consistent with that of the measured rapid quantum capacitance oscillations. Furthermore, we probed weak FP interference patterns in the $Cd_3As_2$ Josephson junction using the RF reflected signal, which is not measurable in DC conductance measurements. This indicates that the enhanced sensitivity of our RF reflectometry technique is powerful for probing quantum transport phenomena in quantum devices. The observation of FP interference in long-channel $Cd_3As_2$ nanowires opens up possibilities for future quantum device applications, particularly in developing topological Andreev and gatemon qubits based on $Cd_3As_2$.

**Supporting Information**

Theoretical model of 1D nanowire for Fabry-Pérot (FP) interference (Supplement Notes S1-S3, Figures S1-S5); full-scale of FP interference 2D plot over entire measurement range (Figure S6); supercurrent in $Cd_3As_2$ nanowire Josephson junctions as a function of channel length (Device C and D, Figure S7) (PDF)

**ACKNOWLEDGEMENTS**

This work is supported by the Mid-Career Researcher Program of NRF (RS-2023-NR076585 and RS-2023-00278511) and the DGIST R&D Program of the Ministry of Science, ICT, and Future Planning (25-ET-02, 25-SENS2-09). This work was supported by Institute for Information & communications Technology Promotion (IITP) (2021-0-01511), and NRF (RS-2023-00269616, RS-2023-00258732, RS-2024-00402302, RS-2025-00557045, RS-2023-00285353) grants funded by the Korean government (MSIT & MOE).

**AUTHOR CONTRIBUTIONS**

S.J.A, J.K., S.-B.S. and H.K. fabricated the device. S.J.A, J.K., and M.J. measured the devices and analyzed the data, with help from J.S., N.M. and C.S., and K.P. and J.P. performed nanowire synthesis. M.-C.J., Z.B.S., N.M. and M.B.A.J performed the numerical simulations. M.J. and J.S. supervised the experiments presented in this paper. S.J.A., J.K., M.-C.J., J.S. and M.J. prepared the paper with input from all authors.

**Notes**

The authors declare no competing financial interest.

**Table of Contents (TOC)**

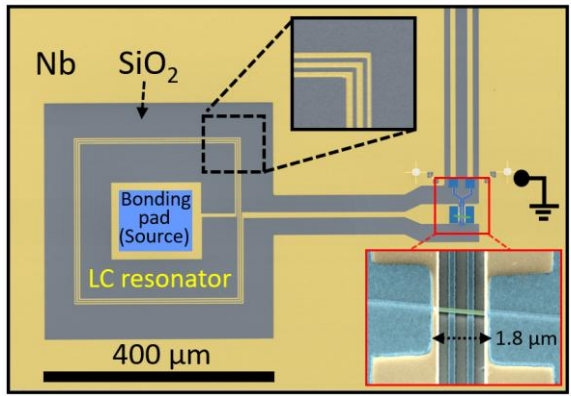 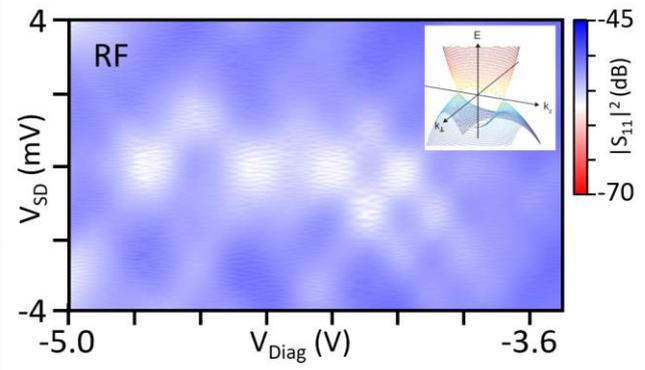

# Supporting Information:
# Radio-Frequency Detection of Fabry-Pérot Interference and Quantum Capacitance in Long-Channel Three-Dimensional Dirac Semimetal Cd$_3$As$_2$ Nanowires

Sung Jin An,[1,¶] Jisu Kim,[1,2,¶] Myung-Chul Jung,[3,4,¶] Kidong Park,[1] Jeunghee Park,[5] Seung-Bo Shim,[6] Hakseong Kim,[6] Zhuo Bin Siu,[7] Mansoor B. A. Jalil,[7] Christian Schönenberger,[8] Nojoon Myoung,[3,4,*] Jungpil Seo,[2,*] and Minkyung Jung[1,9,10,*]

[1] DGIST Research Institute, DGIST, Daegu 42988, Korea

[2] Department of Physics and Chemistry, DGIST, Daegu 42988, Korea

[3] Department of Physics Education, Chosun University, Gwangju 61452, Korea

[4] Institute of Well-Aging Medicate & CSU G-LAMP Project Group, Gwangju 61452, Korea

[5] Department of Advanced Materials Chemistry, Korea University, Sejong 30019, Korea

[6] Quantum Technology Institute, Korea Research Institute of Standards and Science, 34113 Daejeon, South Korea

[7] Electrical and Computer Engineering, National University of Singapore, Singapore 117576, Singapore

[8] Department of Physics, University of Basel, Klingelbergstrasse 82, CH-4056 Basel, Switzerland

[9] Institute of next generation semiconductor technology (INST), DGIST, Daegu 42988, Korea

[10] Department of Interdisciplinary Engineering, DGIST, Daegu 42988, Korea

¶ *S.-J.A, J.K. M.-C.J. contributed equally to this work.*

*Corresponding author. Email: nmyoung@chosun.ac.kr, jseo@dgist.ac.kr, minkyung.jung@dgist.ac.kr

**Supplement Note. S1: Theoretical model of 1D nanowire**

To interpret the Fabry-Pérot (FP) conductance oscillations observed in the experiments, we developed a simple theoretical model based on one-dimensional cylindrical nanowires that replicates the experimental geometry, as described by the Schrödinger equation.

The Schrödinger equation for a nanowire with radius $a$ is

$$-\frac{\hbar^2}{2m^*}\left[\frac{1}{\rho}\frac{\partial}{\partial \rho}\left(\rho \frac{\partial}{\partial \rho}\right) + \frac{1}{\rho^2}\frac{\partial^2}{\partial \phi^2} + \frac{\partial^2}{\partial z^2}\right]\psi = [E - V(z)],$$

where $V(z)$ is the electrostatic potential as a function of $z$.

The general solution of our model is given by

$$\psi(\vec{r}) = J_m(k\rho)e^{im\phi}e^{ik_z z}.$$

Here, $J_m(k\rho)$ is the Bessel function of the first kind, $\rho$ is the radial coordinate, $e^{im\phi}$ is the solution for the azimuthal degree of freedom, and $e^{ik_z z}$ is a plane wave form corresponding to the propagation of the electron along the $z$ direction.

The boundary condition with respect to the $\rho$ coordinate at the boundary of the nanowire

$$\psi(r = a) = 0,$$

gives

$$R(\rho = a) = J_m(ka) = 0.$$

$ka$ is hence a zero of the Bessel function $J_m$, which results in discrete values for $k$, which we index with an integer $n$. The general solution of our system with radius $a$ is therefore

$$\psi(\vec{r}) = J_m(k_{m,n}\rho)e^{im\phi}e^{ik_z z},$$

where $k_{m,n}a$ is the $n$th zero of $J_m$.

By substituting the general solution into the Schrodinger equation, we obtain the energy dispersion relation

$$\epsilon = k_{m,n}^2 + k_z^2 + v(z),$$

where $\epsilon = \frac{2m^*E}{\hbar^2}$ and $v(z) = \frac{2m^*}{\hbar^2}V(z)$.

We consider electron scattering along the nanowire with the back-gate voltage creating a potential barrier in the central region. The one-dimensional barrier problem is divided into three regions (see Supplementary Figure S1): Region I, II, and III, which correspond to $z < 0$, $0 < z < L$, and $z > L$, respectively, where $L$ is the length of the suspended region.

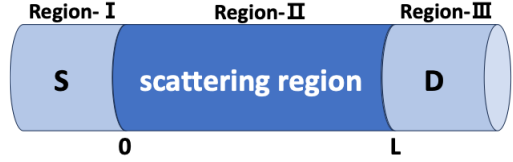

**Figure S1**. Schematic illustration of the nanowire model, which is divided into three regions: Region I (source, S), Region II (scattering region), and Region III (drain, D). A potential is present only in the scattering region (Region II), while the source and drain regions are assumed to have zero potential.

The wavefunctions in each region are:

$$\psi_I = J_0(k_{m,n}\rho)e^{ik_{z;m,n}z} + \sum_{m',n'} r_{m',n'} J_0(k_{m',n'}\rho)e^{-ik_{z;m,n}z},$$

$$\psi_{II} = \sum_{m',n'}[a_{m',n'} J_0(k_{m',n'}\rho)e^{iq_{z;m',n'}z} + b_n J_0(k_n\rho)e^{-iq_{z;m',n'}z}],$$

$$\psi_{III} = \sum_{m',n'} t_{m',n'} J_0(k_{m',n'}\rho)e^{ik_{z;m',n'}z},$$

Note that the longitudinal momenta $k_z$ and $q_z$ also depend on the indices $m$ and $n$ through the following relations:

$$k_{z;m,n} = \sqrt{\epsilon - k_{m,n}^2}, \quad q_z = \sqrt{[\epsilon - v_0] - k_{m,n}^2}.$$

By applying the boundary condition of continuity of wavefunction and its derivatives at $z = 0$ and $z = L$ and exploiting the orthogonality relations $\int_{-\pi}^{\pi} \exp(i(m - m')\phi)\, d\phi \propto \delta_{m,m'}$ and $\int_0^a \rho J_m(k_{m,n}\rho) J_m(k_{m,n'}\rho)\, d\rho \propto \delta_{n,n'}$, we obtain, for each incident $k_{m,n}$ mode, a $4 \times 4$ matrix equation (we omit the m, n subscripts in $k_z$ and $q_z$ for notational simplicity)

$$\begin{pmatrix} -1 & 1 & 1 & 0 \\ k_z & q_z & -q_z & 0 \\ 0 & -e^{iq_zL} & e^{-iq_zL} & e^{ik_zL} \\ 0 & q_z e^{iq_zL} & -q_z e^{-iq_zL} & -k_z e^{ik_zL} \end{pmatrix} \begin{pmatrix} r_{m,n} \\ a_{m,n} \\ b_{m,n} \\ t_{m,n} \end{pmatrix} = \begin{pmatrix} \Theta(\epsilon - \epsilon_{m,n}) \\ k_z \Theta(\epsilon - \epsilon_{m,n}) \\ 0 \\ 0 \end{pmatrix}.$$

The term $\Theta(\epsilon - \epsilon_{m,n})$ indicates that the scattering problem for the $(m,n)$th subband is valid only when the electron energy exceeds the subband edge $\epsilon_{m,n} = k_{m,n}^2$.

Using the Landauer–Büttiker formalism, the linear-response conductance is

$$G = \frac{2e^2}{h} \sum_n |t_n|^2.$$

Figure 3(d) in the main text models the conductance with gate voltage (varying the potential barrier height) for fixed incident energy near the third subband onset. Quantum interference arises due to the superposition of oscillatory contributions from different subbands.

**Supplement Note. S2: Theoretical result of quantum conductance simulations**

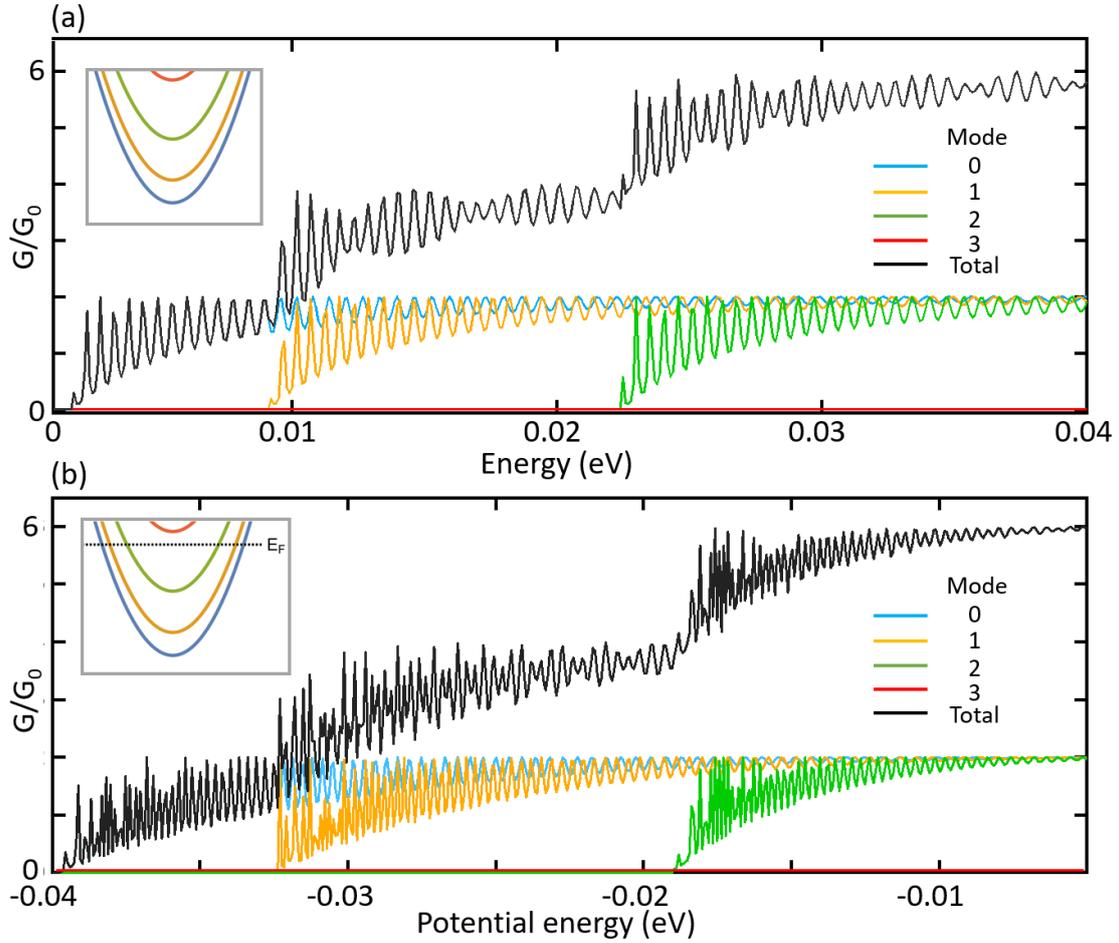

**Figure S2**. (a) Quantum conductance as a function of electron energy, calculated from mode-resolved transmission probabilities. Colored solid lines represent the transmission of individual transverse modes (Modes 0–3), while the black line indicates the total conductance summed over all modes. (Inset) Schematic band structure of the nanowire, where each parabolic curve corresponds to a discrete transverse mode. (b) Quantum conductance as a function of gate-induced potential energy. Colored solid lines denote the transmission of individual transverse modes (Modes 0–3), and the black line represents the total conductance. (Inset) Schematic band structure of the nanowire with a fixed Fermi energy $E_F$ = 0.0415 eV (dashed line), indicating three active propagating modes corresponding to the intersections of $E_F$ with the subband dispersions.

Figure S2 presents the calculated quantum conductance of a cylindrical 1D nanowire under a fixed potential barrier, based on the model introduced in Supplementary Note S1. Mode-resolved transmissions for four transverse modes (Modes 0–3) are shown as colored lines in Figure S2(a), while the total conductance, obtained by summing over all occupied modes, is shown in black. The total conductance exhibits a stepwise increase as a function of electron energy, consistent with quantized transport in 1D systems. Superimposed on this stepwise behavior are pronounced oscillations that reflect multimode FP interference. In the third conductance plateau—where three subbands are occupied—the interference between multiple modes results in beating patterns, consistent with the superposition of oscillations from distinct wavevectors.

To more closely emulate experimental conditions, we fixed the Fermi energy between the third and fourth subbands [see inset of Figure S2(b)] and varied the electrostatic potential in the suspended region via the gate voltage. In this regime, three modes contribute to transport, with their transverse momenta and transmission phases varying with the local potential. The resulting conductance, shown in Figure S2(b), exhibits clear mode-dependent interference patterns and a gate-tunable transition between different subband configurations. The simulated conductance captures both the stepwise quantization and the multimode interference observed experimentally, supporting the interpretation that the aperiodic FP oscillations originate from coherent transport through multiple transverse modes in the $Cd_3As_2$ nanowire.

**Supplement Note. S3: Theoretical analysis of quantum conductance oscillation**

Figure S3 shows the mode-resolved quantum conductance of a cylindrical nanowire as a function of energy, highlighting FP oscillations for the lowest three transverse modes (Modes 0–2). The energy spacing ($\Delta E_i$) between adjacent conductance peaks is extracted to quantify the interference characteristics of each mode. The conductance oscillations for each mode exhibit nearly constant spacing, indicating well-defined resonance conditions. The average energy spacings are $\Delta E \approx 0.81$ meV (Mode 0), 0.75 meV (Mode 1), and 0.62 meV (Mode 2), corresponding to dominant FP oscillation frequencies of 1.23, 1.33, and 1.61 meV$^{-1}$, respectively. These results confirm that each mode sustains coherent transport with distinct phase accumulation, leading to superimposed oscillations in the total conductance.

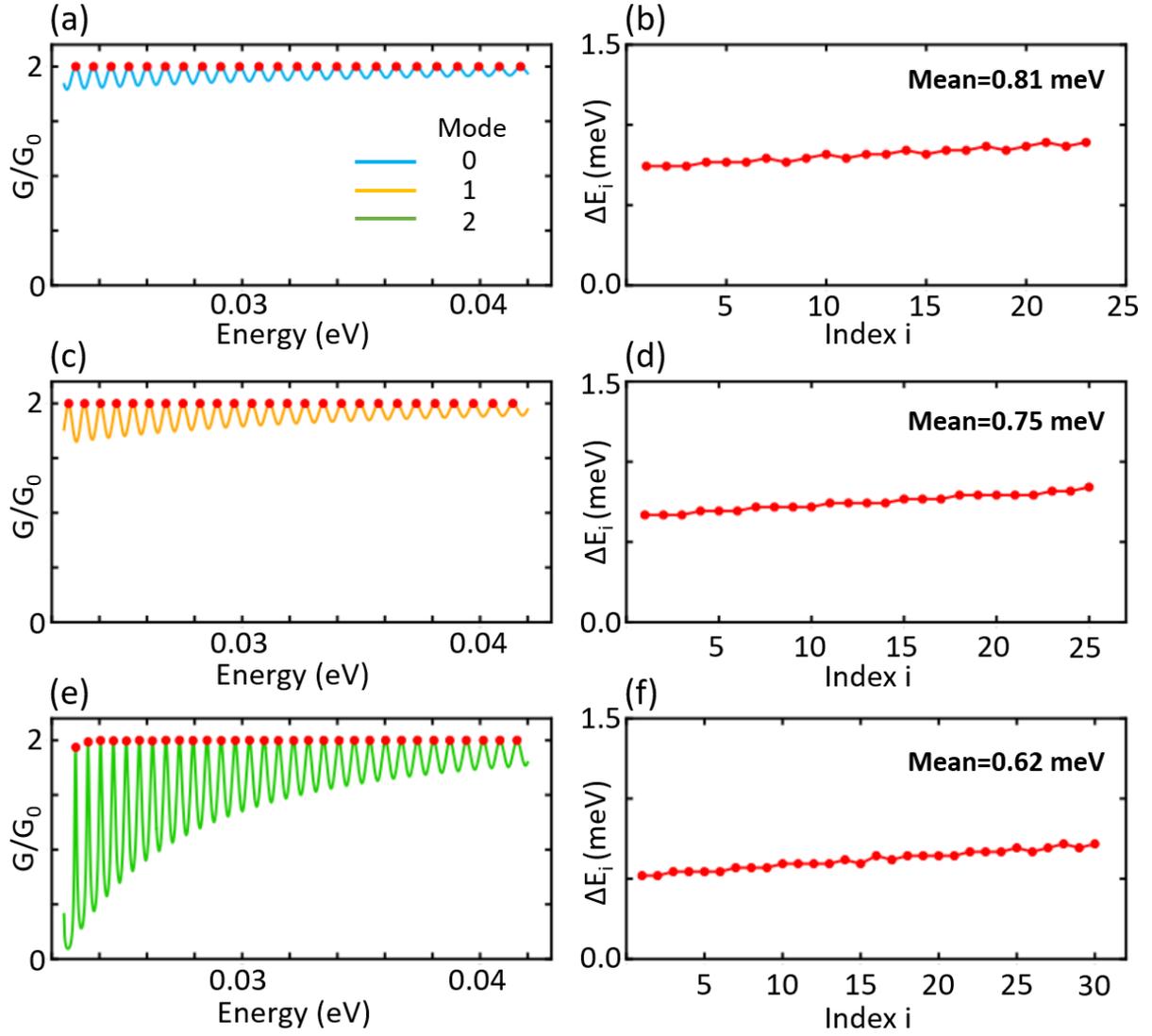

**Figure S3.** (a,c,e) Mode-resolved quantum conductance $G/G_0$ versus energy for Modes 0–2, highlighting Fabry–Pérot oscillations. Red circles indicate identified conductance peaks. (b,d,f) Energy spacing $\Delta E_i$ between adjacent peaks for each mode. Mean values are indicated in each panel.

To further investigate the gate-voltage dependence of multimode interference, we analyzed the conductance as a function of gate-induced potential energy, shown in Figure S4. The average energy spacings extracted from Figure S4 are 0.62 meV (Mode 0), 0.54 meV (Mode 1), and 0.36 meV (Mode 2), corresponding to frequencies of 1.63, 1.85, and 2.78 meV$^{-1}$, respectively. To estimate the dominant frequency of FP oscillations for each mode, we used the inverse of the mean energy spacing between adjacent peaks, as shown in Figures S3(b,d,f) and S4(b,d,f).

The frequencies extracted from both energy- and gate-potential-dependent simulations are in close agreement and consistently account for the densely spaced oscillation features observed experimentally in Figure 3(c).

Figure S5(a) presents the total quantum conductance from three occupied subbands, around $G/G_0 \approx 6$, as a function of gate-induced potential energy. In contrast to the single-mode regime, the conductance shows irregular oscillation spacings arising from the superposition of multiple interference modes. This behavior is consistent with the experimentally observed beating patterns and aperiodic oscillations discussed in the main text.

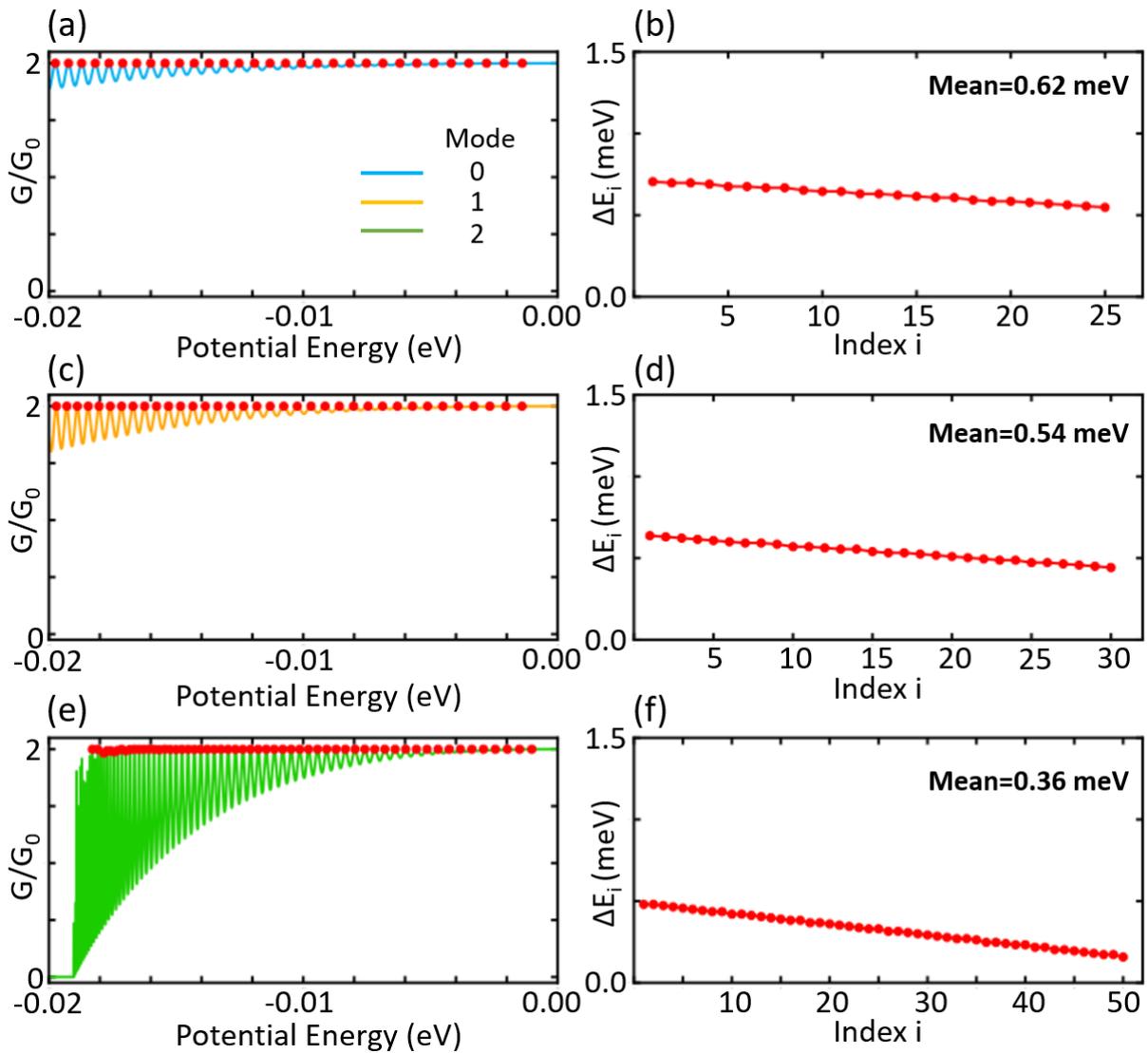

**Figure S4.** (a, c, e) Mode-resolved quantum conductance $G/G_0$ as a function of gate-induced potential energy for Modes 0–2. Red markers indicate Fabry–Pérot oscillation peaks. (b, d, f)

Energy spacing $\Delta E_i$ between adjacent peaks for each mode. Mean values are indicated in each panel.

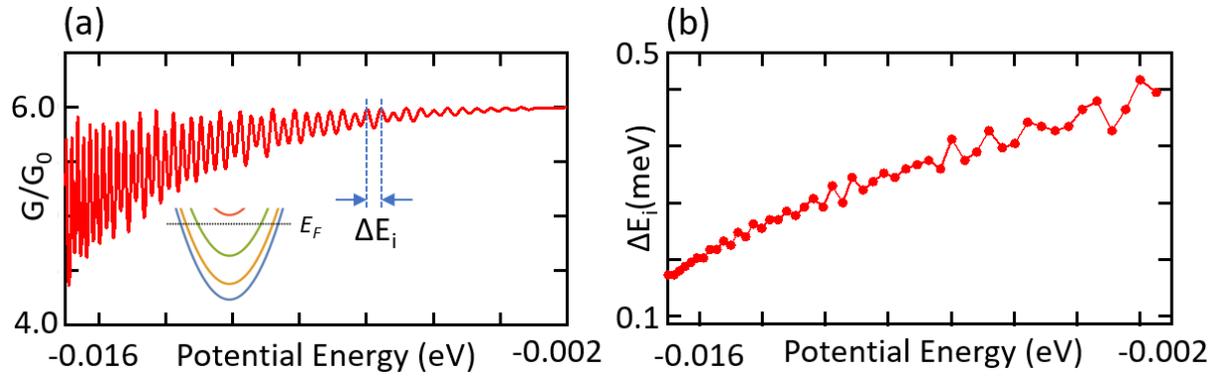

**Figure S5.** (a) Total quantum conductance $G/G_0$ as a function of gate-induced potential energy for the third subband. (b) Energy difference ($\Delta E_i$) between peaks as a function of gate-induced potential energy.

**The full-scale of 2D plot over the entire measurement range**

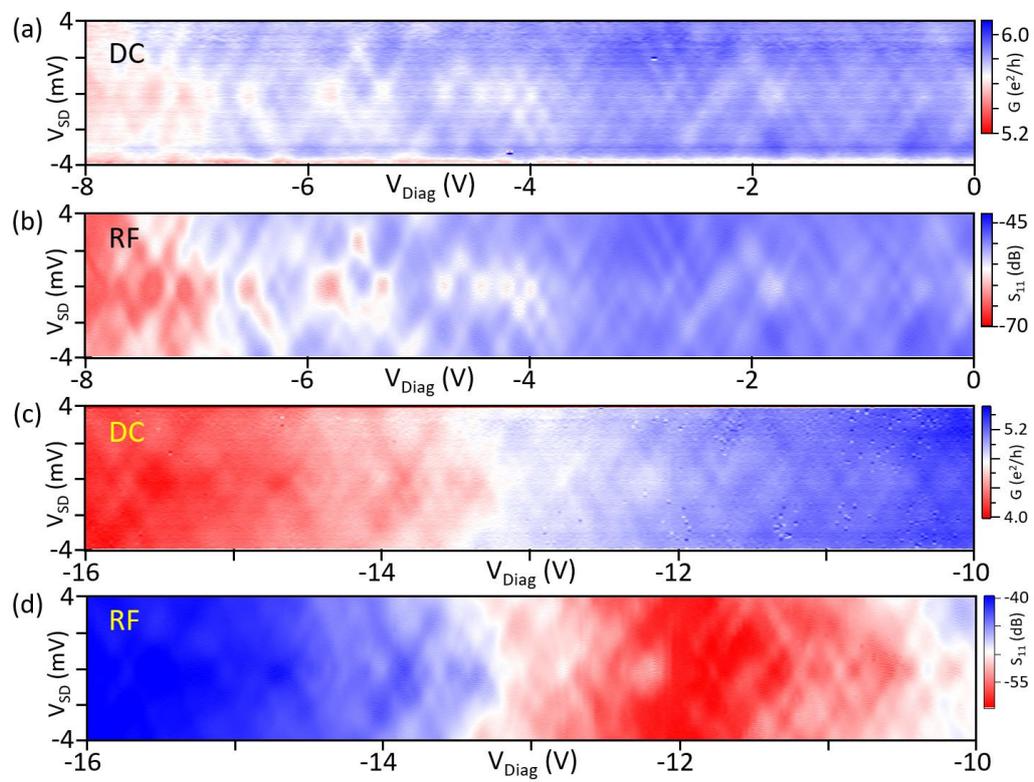

**Figure S6.** (a,c) Differential conductance and (b,d) RF reflection coefficient ($|S_{11}|^2$) measured in long gate voltage ranges.

**Supercurrent in Cd$_3$As$_2$ nanowire Josephson junctions depending on channel length (Device C and D)**

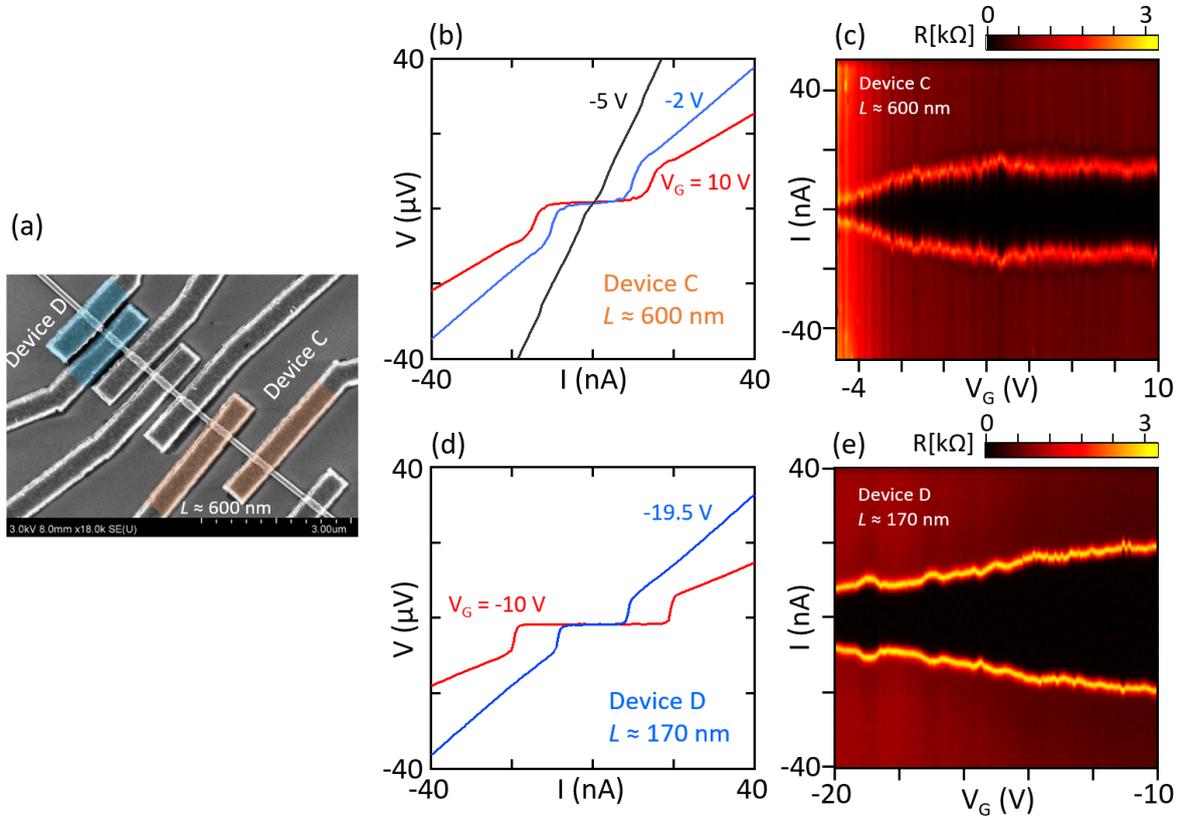

**Figure S7.** (a) Cd$_3$As$_2$ nanowire Josephson junctions with different channel length. (b) *I-V* characteristics at different gate voltages for Device C. (c) Differential resistance *dV/dI* as a function of current bias *I* and gate voltage $V_G$ for Device C. (d) *I-V* characteristics at different gate voltages for Device D. (e) Differential resistance *dV/dI* as a function of current bias *I* and gate voltage $V_G$ for Device D.